\documentstyle[12pt,epsfig]{article}

\newcommand{\be}{\begin{equation}}
\newcommand{\ee}{\end{equation}}
\newcommand{\bea}{\begin{eqnarray}}
\newcommand{\eea}{\end{eqnarray}}
\newcommand{\beast}{\begin{eqnarray*}}
\newcommand{\eeast}{\end{eqnarray*}}
\newcommand{\ds}{\displaystyle}
\begin{document}
\begin{titlepage}
\begin{flushright}
DO-TH-94/17\\
July 1994
\end{flushright}

\vspace{20mm}
\begin{center}
{\Large \bf
One-loop corrections to the instanton transition
in the two-dimensional Abelian Higgs model }

\vspace{10mm}

{\large  J. Baacke\footnote{e-mail~~
baacke@het.physik.uni-dortmund.de}
and T. Daiber} \\
\vspace{15mm}

{\large Institut f\"ur Physik, Universit\"at Dortmund} \\
{\large D - 44221 Dortmund , Germany}
\vspace{25mm}

\bf{Abstract}
\end{center}
We present an evaluation of the fluctuation determinant
which appears as a prefactor in
the instanton transition rate
for the two-dimensional Abelian Higgs model. The corrections
are found to change the rate at most by a factor of $2$ for
$0.4 < M_H/M_W < 2.0 $.  
\end{titlepage}


\section{Introduction}
\par

The Abelian Higgs model in $(1+1)$ has found considerable
attention recently
since on the one hand it shares certain features with the
electroweak theory and on the other hand it is simple enough
to serve as a theoretical and numerical laboratory.

In the context of the baryon number violation 
the high temperature sphaleron transition has been studied
\cite{GriRuSha}-\cite{SmiTa}, for which exact classical 
solutions and an exact expression of the sphaleron determinant 
\cite{BoSha}-\cite{Bo} are known, providing thus a complete 
one-loop semiclassical
transition rate which can be studied numerically 
on the lattice, e.g. by
measuring the fluctuations of the Chern-Simons number.

Another prominent feature of the model is the existence
of instanton solutions \cite{NiOl} that give rise
again to fluctuations in
the topological charge of the vacuum and thereby to baryon number
violation. It has also been used \cite{KriRi} to study the
possibility of baryon number violation in high
energy scattering processes. 

If the parameters of the model are chosen
appropriately, the instantons are suffifiently rare and
it should be sufficient to consider a dilute gas of
instantons with Chern-Simons charge $q=\pm 1$.
For such a dilute gas of instantons the
transition rate, or equivalently the density of 
instantons in the Euclidean plane, is given by
\cite{Co}
\be                                         \label{rate}
\Gamma =
 \frac{S(\phi_{cl})}{2\pi } 
{\cal D}^{-1/2} \exp (-S(\phi_{cl})-S_{ct}(\phi_{cl}))
\end{equation}
to 1-loop accuracy. Here $S(\phi_{cl})$ is the 
instanton action. The coefficient ${\cal D}$
represents the effect of quantum fluctuations around the instanton
configuration and arises from the Gaussian approximation to
the functional integral. This is the object
whose computation we will consider here.
It is given in general form by
\be             \label{det}
{\cal D}
=\frac{\det'( {\cal M})  }{\det( {\cal M}^0 )}
=\exp(2 S_{eff}^{1-loop})
\ee
where the second equation relates it to the
one-loop effective action.
The operators ${\cal M}$ are the fluctuation operators obtained
by taking the second functional derivative of the action
at the instanton and vacuum background field configurations.
The prime in the determinant implies omitting of the two translation
zero modes. The first prefactor $S(\phi_{cl})/2\pi$ takes
into account of the integration of the 
translation mode collective coordinates. 
Finally the counterterm action $S_{ct}$ in the exponent will
absorb the ultraviolet divergences of ${\cal D}$.
One may also include
a corresponding determinant for fermions, which for massless
fermions is even known analytically \cite{NiSch} (see below).
However in lattice simulations the instanton rate and
therefore fermion number violation can be measured
by studying fluctuations of the Chern-Simons number and
it is not necessary to include fermions. 

For $M_H/M_W \neq 1$ even the classical instanton
profiles are known only numerically, so the evaluation of
the effective action has to be performed numerically.
A method for such computations has been proposed
previously \cite{Baaeff}; it has been used recently for 
the computation of the fluctuation determinant
of the electroweak sphaleron \cite{BaaJu} and 
for the case considered here in 
\cite{Dai}, on which the present work is based.

This paper is organized as follows:
In the next section we outline the basic relations of the
Abelian Higgs model. The fluctuation operator is derived in
section 3, its partial wave reduction in section 4.
The method of computation is presented in section 5.
In section 6 we consider the renormalization of the
effective action and the removal of zero modes.
The results are presented and discussed in section 7.

\setcounter{equation}{0}
\section{Basic relations}
\par

The Abelian Higgs model in (1+1) dimensions is
defined by the Lagrange density (written in the Euclidean form
relevant here)
\be
{\cal L}=\frac{1}{4}(F_{\mu\nu})^2
+\frac{1}{2}|D_\mu\phi|^2+\frac{\lambda}
{4}\left(|\phi|^2-v^2\right)^2+{\cal L}_f^{} \; .  
\ee

Here
\beast
F_{\mu\nu}&=&\partial_\mu A_\nu-\partial_\nu A_\mu \\
D_\mu&=&\partial_\mu-igA_\mu \\ 
{\cal L}_f^{}&=&i\sum_{i=1}^{n_f}\overline
\Psi_L^{\left(i\right)}\hat{D}_L^{}
\Psi_L^{\left(i\right)}+i\sum_{j=1}^{n_f}
\overline\Psi_R^{\left(j\right)}
\hat{D}_R^{}\Psi_R^{\left(j\right)} \\ 
\hat{D}_{L(R)}^{}&=&\gamma_\mu\left(\partial_\mu\mp igA_\mu\right)\ .
\eeast

The particle spectrum consists of Higgs bosons of mass
$m_H^2=2\lambda v^2$, vector bosons of mass $m_W^2=g^2v^2$ and
left (right) handed massless fermions of charge g.
The anomaly of the gauge invariant fermionic current
\be
J_\mu=\sum_{i=1}^{n_f}\overline\Psi_L^{(i)}\gamma_\mu
\Psi_L^{(i)}+\sum_{j=1}^{n_f}
\overline\Psi_R^{(j)}\gamma_\mu\Psi_R^{(j)} 
\ee
is given by
\be
\partial_\mu J_\mu=2n_f\left(\frac{g}{4\pi}\varepsilon_{\mu\nu}
F_{\mu\nu}\right) \; .   
\ee

The integral over the divergence of the current which measures
the baryon number violation is given by
\bea
\Delta F&=&\int d^2x\,\partial_\mu J_\mu  \\
&=&2n_f \left( \frac{g}{4\pi}\int d^2x \, 
\varepsilon_{\mu\nu}F_{\mu\nu} \right) \equiv 2n_fq \; ,
\eea

Here $q$ denotes the  Chern-Simons term in
two dimensions (siehe z.B. \cite{Ja}) and
baryon number violation is therefore related to
euclidean gauge field configurations with nonvanishing topological
charge $q$. These are the instanton solutions
which mediate tunneling transitions changing the
topological charge by $q$ units.
We will assume here that the instantons
transitions are described sufficiently well by a
dilute gas of instantons with Chern Simons
number $q=\pm 1$, a situation for which the rate
formula given in the introduction is supposed to hold.

A structure that exhibits such a topological charge and satisfies
the euclidean equations of motion is given by the
Nielsen-Olesen vortex \cite{NiOl}. The spherically symmetric
ansatz for this solution is given by
\begin{eqnarray}
A_\mu^{cl}(x)&=&\frac{\varepsilon_{\mu\nu}x_\nu}{gr^2}A(r) \\
\phi^{cl}(x)&=&vf(r)e^{i\varphi(x)} \; .
\end{eqnarray}

In order to have a purely real Higgs field one performs a
gauge transformation
\bea \label{gaugetr}
\phi &\to& e^{-i\varphi}\phi\ \nonumber \\
 A_\mu&\to& A_\mu-\partial_\mu\varphi/g \nonumber \\
 \Psi_{L(R)}&\to& e^{\mp i\varphi}\Psi_{L(R)}
\eea
to obtain the instanton fields in the singular gauge
\bea
A^{cl}_\mu (x)&=&\frac{\varepsilon_{\mu\nu}x_\nu}{gr^2}(A(r)+1)  \\
\phi^{cl}(x)&=&vf(r)
\eea

With this ansatz the euclidean action takes the form
\bea
S_{cl}&=&\pi v^2
 \int^{\infty}_{0}\!\!\! dr \left(\frac{1}{rm_W^2}\left(
\frac{dA(r)}{dr}\right)^2\!\!+r\left(\frac{df(r)}{dr}\right)^2\!\!
+\frac{f^2(r)}{r}\left(A(r)+1\right)^2\!\! \right. \nonumber \\ 
&+&\left. \frac{rm^2_H}{4}\left(
f^2(r)-1\right)^2\!\right)
\eea

For the case $M_H=M_W$ an exact solution to the
variational equation is known \cite{deVSha} for which
the classical action takes the value $S_{cl}=\pi v^2$.
We will consider here the general case, however, for which
the classical equations of motion
\bea
\left(\frac{\partial^2}{\partial r^2}+
\frac{1}{r}\frac{\partial}{\partial r}
-\frac{\left(A(r)+1\right)^2}{r^2}-
\frac{m^2_H}{2}\left(f^2(r)-1\right)
\right)f(r)&=&0 \\\left(\frac{\partial^2}
{\partial r^2}-\frac{1}{r}
\frac{\partial}{\partial r}-m^2_W f^2(r)\right)
A(r)&=&m^2_W f^2(r) \ .
\eea
have to be solved numerically.
 
Imposing the boundary conditions on the profile functions
\be  \label{rb}
\begin{array}{rcccccr}
A(r)&\stackrel{\scriptscriptstyle{r\to 0}}
{\longrightarrow}& const\cdot r^2 
&,&A(r)&\stackrel{\scriptscriptstyle{r\to\infty}}
{\longrightarrow}&-1 \\ 
f(r)&\stackrel{\scriptscriptstyle{r\to 0}}
{\longrightarrow}&const\cdot r 
&,&f(r)&\stackrel{\scriptscriptstyle{r\to\infty}}
{\longrightarrow}&1 
\end{array}
\ee
the Chern-Simons number is $1$ and the action is finite.

Since we will consider fluctuations around these solutions later
on a good numerical accuracy for the profile functions
$f(r)$ and $A(r)$ is required. We have found that the
method used previously by Bais and Primack \cite{BaiPr}
in order to obtain precise profiles for the 't Hooft-Polyakov
monopole is very suitable also in this context.
The values for the classical action -  which determine also
the translation mode prefactor - are
given in Table 1.  They agree almost perfectly with
with the results of Jacobs and Rebbi \cite{JaRe}.

The classical action in units of $\pi v^2$
is plotted in Fig. 1 for $.4 < M_H/M_W < 2$. The plot
suggests a power law behaviour $(M_H/M_W)^\rho$ where
$0.40 < \rho < .43$; our data and those of Ref. \cite{JaRe}
are precise enough to rule
out an exact power dependence.
While the best fit is
obtained with $\rho \simeq .41$ a suggestive number
in this range
is $\rho=1-\gamma$, where $\gamma$ is Eulers constant.
This could be an asymptotic dependence for large
$M_H/M_W$, it is also displayed in Fig.1.

Though we are not interested here in the effect of fermionic
fluctuations, we could not resist to use our profiles to
calculate the fermion determinant for massless fermions.
It is known exactly \cite{NiSch}; expressed as an
effective action it is, per (left plus right handed) fermion
and with fermionic zero modes removed,
\be \label{fermac}
S_{eff}^{ferm} = -\frac{1}{2\pi} \int d^2x \alpha \partial^2
\alpha
\ee
where for the instanton
\be
\alpha(r) = \int_0^r dr \frac{A(r)}{r}
\ee
The results are given in Table 1 and plotted with the
bosonic effective action in Fig. 4.

\setcounter{equation}{0}
\section{The fluctuation operator}
The fluctuation operator is
defined in general form as
\be
{\cal M} =  \frac{\delta^2S}{\delta \psi^*_i (x) \delta \psi_j (x')}
|_{\psi_k=\psi_k^{cl}}
\ee
where $\psi_i$ denote the fluctuating fields and
$\psi_i^{cl}$ the ``classical'' background field
configuration which here
will be the instanton and the vacuum configurations. If the fields
are expanded around the background field as
$\psi_i = \psi_i^{cl} + \phi_i$
and the Langrange density is expanded accordingly then the
fluctuation operator is related to the second order Lagrange density
via
\be
{\cal L}^{II} =  \frac{1}{2}  \phi^*_i {\cal M}_{ij}  \phi_j
\ee

In terms of the fluctuation operators
${\cal M}$ on the instanton  and ${\cal M}^0$ on the
vacuum backgrounds the effective action
is defined as
\be
S_{eff} = \frac{1}{2} \ln \left\{ \frac{\det' {\cal M}}
{\det {\cal M}^0} \right\}
\ee

For our specific model we expand as
\bea
A_\mu&=&A_\mu^{cl}+a_\mu  \\ 
\phi&=&\phi^{cl}+\varphi \ ,
\eea

In order to eliminate the gauge degrees of
freedom we introduce, as in Ref. \cite{KriRi},
 the background gauge function
\be
{\cal F}(A)=\partial_\mu
A_\mu+\frac{ig}{2}\left((\phi^{ cl})^\ast\phi-
\phi^{cl}\phi^\ast\right)  
\ee
which leads in the Feynman background gauge to the gauge fixing
Lagrange density
\bea
{\cal L}_{GF}^{I\hspace{-.05cm}I}&=
&\left(\frac{1}{2}{\cal F}^2(A)\right)
^{I\hspace{-.05cm}I} \nonumber \\ 
&=&\frac{1}{2}(\partial_\mu a_\mu)^2
-\frac{ig}{2}a_\mu(\varphi\partial_\mu\phi^{cl}
+\phi^{cl}\partial_\mu\varphi
-\varphi^\ast\partial_\mu\phi^{cl}-\phi^{cl}
\partial_\mu\varphi^\ast) \\
&&-\frac{g^2}{8}(\phi^{cl})^2(\varphi-\varphi^\ast)^2  \ ,
\nonumber \eea
The corresponding Fadeev-Popov Lagrangian becomes
\be
{\cal L}_{FP} =\frac{1}{2} \eta^\ast(-\partial^2+g^2
\left(\phi^{cl}\right)^2)\eta
\ee
In terms of the real components $\varphi= \varphi_1+i\varphi_2$
and $\eta = (\eta_1 + i \eta_2)/\sqrt{2}$ the second order Lagrange
density becomes now
\bea \label{lag2}
\left({\cal L}+{\cal L}_{GF}
+{\cal L}_{FP}\right)^{I\hspace{-0.05cm}I}&=&
a_\mu\frac{1}{2}\left(-\partial^2+g^2\phi^2\right)a_\mu \nonumber \\
&&+\varphi_1\frac{1}{2}
\left(-\partial^2+g^2A_\mu^2+\lambda\left(3\phi^2-
v^2\right)\right)\varphi_1  \nonumber \\  
&&+\,\varphi_2\frac{1}{2}\left(-\partial^2+g^2A_\mu^2+g^2\phi^2+
\lambda\left(\phi^2-v^2\right)\right)\varphi_2 \nonumber \\
&&+\,\varphi_2(gA_\mu\partial_\mu)\varphi_1
+\varphi_1(-gA_\mu\partial_\mu)\varphi_2  \\
&&+\,a_\mu(2g^2A_\mu\phi)\varphi_1
+\,a_\mu(2g\partial_\mu\phi)\varphi_2 \nonumber \\
&&+\eta_1\frac{1}{2}\left(-\partial^2+g^2\phi^2\right)\eta_1
+\eta_2\frac{1}{2}\left(-\partial^2+g^2\phi^2\right)\eta_2
\nonumber  
\eea
Specifying now the fluctuating fields
$ (\phi_1,\phi_2,\phi_3,\phi_4,\phi_5)$ as
$ (a_1,a_2,\varphi_1,\varphi_2,\eta_{12})$
the nonvanishing components of ${\cal M}$ are

\bigskip
$\begin{array}{rcl@{\qquad}rcl}
{\cal M}_{11}&=&\ds -\partial^2 + g^2 \phi^2&
{\cal M}_{22}&=&\ds -\partial^2 + g^2 \phi^2 \\
{\cal M}_{13}&=& 2 g^2
A_1 \phi&{\cal M}_{14}&=& 2 g \partial_1 \phi \\
{\cal M}_{23}&=&2 g^2
A_2 \phi&{\cal M}_{24}&=& 2 g \partial_2 \phi \\
{\cal M}_{33}&=&\ds -\partial^2+g^2A_\mu^2 +\lambda(3\phi^2-v^2)&
{\cal M}_{34}&=&\ds -g A_\mu \partial_\mu  \\
{\cal M}_{44}&=&\ds -\partial^2 +g^2 A_\mu^2 + g^2 \phi^2
+\lambda (\phi^2-v^2)& {\cal M}_{43}&=& g A_\mu \partial_\mu \\
{\cal M}_{55}&=&-\partial^2+g^2\left(\phi^{cl}\right)^2&
&&
\end{array}$ 
\bigskip

It is understood that the contribution of the Fadeev-Popov
operator ${\cal M}_{55}$ enters with a negative sign and a factor 
$2$ into the definition of the effective action.
The fluctuation operators for the instanton and vacuum
background are now obtained by substituting the corresponding
classical fields.
The vacuum fluctuation operator becomes a diagonal
matrix of Klein-Gordon operators with masses
$(M_W,M_W,M_W,M_H,M_W)$. It is convenient
to introduce a potential $\cal V$ via
\be
{\cal M} = {\cal M}^0 + {\cal V}
\ee
The potential ${\cal V}$ will be specified below after partial
wave decomposition.

\setcounter{equation}{0}
\section{Partial wave decomposition}
\par

The fluctuation operator ${\cal M}$ can be decomposed 
into partial waves and its the determinant decomposes
accordingly.
\be
\ln \det {\cal M} = \sum_{n=-\infty}^{+\infty}
\ln \det {\bf M}_n
\ee
We introduce the following partial wave decomposition for
fields 
\beast
\vec{a}&=&\sum_{n=-\infty}^{+\infty}
b_n(r)\left(\begin{array}{c}\cos\varphi\\\sin\varphi
\end{array}\right)\frac{e^{in\varphi}}{\sqrt{2\pi}}+ic_n(r)\left(
\begin{array}{c}-\sin\varphi\\\cos\varphi\end{array}\right)
\frac{e^{in\varphi}}{\sqrt{2\pi}}  \\
\varphi_1&=&\sum_{n=-\infty}^{+\infty}
h_n(r)\frac{e^{in\varphi}}{\sqrt{2\pi}}  \\
\varphi_2&=&\sum_{n=-\infty}^{+\infty}
\tilde{h}_n(r)\frac{e^{in\varphi}}{\sqrt{2\pi}}  \\
\eta_{12}&=
&\sum_{n=-\infty}^{+\infty} g_n(r)\frac{e^{in\varphi}}{\sqrt{2\pi}}
\eeast  

After inserting these expressions into the
Lagrange density and using the reality conditions for
the fields one finds that the following combinations are
relatively real and make the fluctuation operators
symmetric:
\beast
F^n_1(r)&=&\frac{1}{2}(b_n(r)+c_n(r))  \\
F^n_2(r)&=&\frac{1}{2}(b_n(r)-c_n(r))  \\
F^n_3(r)&=&\tilde{h}_n(r)  \\
F^n_4(r)&=&ih_n(r)  \\
F^n_5(r)&=&g_n(r)
\eeast
Writing the partial fluctuation operators - omitting the index
$n$ in the following - as
\be
{\bf M} = {\bf M}^0 + {\bf V}
\ee
the free operators ${\bf M}^0$ become diagonal matrices with
elements
\be
M^0_{ii}= -\frac{d^2}{dr^2}-\frac{1}{r}\frac{d}{dr}
+\frac{n_i^2}{r^2} + M_i^2
\ee
where $(n_i) = (n-1,n+1,n,n,n)$ and $(M_i)=(M_W,M_W,M_W,M_H,M_W)$.
The potentials ${\bf V}$ takes the elements

\bigskip
$\begin{array}{rcl@{\qquad}rcl}
{\bf V}_{11}^n&=&m_W^2\left(f^2-1\right) &{\bf V}_{12}^n&=&0\\
{\bf V}_{13}^n&=&\sqrt{2}m_W^{}f'&{\bf V}_{14}^n&=&\ds\sqrt{2}m_W^{}f
\frac{A+1}{r}\\
{\bf V}_{22}^n&=&{\bf V}_{11}^n&{\bf V}_{23}^n&=&{\bf V}_{13}^n\\
{\bf V}_{24}^n&=&-{\bf V}_{14}^n&
{\bf V}_{33}^n&=&\ds\frac{(A+1)^2}{r^2}+
\frac{m_H^2}{2}\left(f^2-1\right)
+m_W^2\left(f^2-1\right)\\
{\bf V}_{34}^n&=&\ds -2\frac{A+1}{r^2}n&
{\bf V}_{44}^n&=&\ds\frac{(A+1)^2}
{r^2}+\frac{3}{2}m_H^2\left(f^2-1\right)  \\
{\bf V}_{55}^n&=&m_W^2\left(f^2-1\right)&{\bf V}_{i5}&=&0
\end{array}$
\bigskip \\
Chosing the dimensionless variable $M_Wr$ one realizes that
the fluctuation operator depends only on the ratio
$M_H/M_W$ up to an overall factor $M_W^2$ which cancels in
the ratio with the free operator.

\setcounter{equation}{0}
\section{Computation of the fluctuation determinant}
\par

The method for computing the fluctuation determinant used here
is based on the use uf the Euclidean Green function of the
fluctuation operator. This Green function is defined by
\be
({\cal M} + \nu^2) {\cal G}(\vec x,\vec x',\nu) =
 {\bf 1}\delta(\vec x -\vec x').
\ee
and similarly for the operator ${\cal M}^0$.
It contains the information on
the eigenvalues $\lambda_\alpha$ of the
fluctuation operator via
\be
\int d^2x {\rm Tr} {\cal G}(\vec x,\vec x, \nu) =
\sum_\alpha \frac{1}{\lambda_\alpha^2 + \nu^2}.
\ee
If we define the function $F(\nu)$ via
\be
F(\nu) = \int d^2x {\rm Tr} ({\cal G}(\vec x,\vec x,\nu)-
{\cal G}^0(\vec x,\vec x, \nu))
\ee
we have
\be
-\int_\epsilon^\Lambda d\nu \nu F(\nu) =
\sum_\alpha\frac{1}{2} \ln
\left\{ \frac{(\lambda_\alpha^2+\epsilon^2)
(\lambda_\alpha^{0~2}+\Lambda^2)}{(\lambda_\alpha^{0~2}+
\epsilon^2)(\lambda_\alpha^2+\Lambda^2)} \right\}
\ee
For $\epsilon \to 0$ this is just the logarithm of the ratio of
fluctuation determinants, i. e. the one loop effective
action, regularized with a Pauli-Villars
cutoff. The regularization can be removed, the integral can be taken
to infinity, after subtracting the one loop counterterm
action (see below). Before taking the limit $\epsilon \to 0$
the two zero eigenvalues have to be removed by subtracting
their contribution $ \ln \epsilon^2$. Of course $\epsilon$ has
the dimension of energy. We used here the scale $M_W$
throughout, i.e. by making the radial variable dimensionless.
So $S_{eff}$ contains now a term $-\ln M_W^2$, the numerical
prefactor ${\cal D}^{-1/2}$ and therefore the
rate are computed in units of $M_W^2$. 

After these more formal considerations we have to present a
practical way for computing $F(\nu)$. This is done by using the
partial wave decomposition to write
\be          \label{fnudef}
F(\nu) = \sum_{n=-\infty}^{+\infty} F_n(\nu)
\ee
where
\be \label{fnnudef}
F_n(\nu) = \int dr r {\rm Tr}({\bf G}_n(r,r,\nu)-
{\bf G}_n^0(r,r,\nu)),
\ee
the partial wave Green functions being defined by
\be \label{greendgl}
({\bf M}_n +\nu^2) {\bf G}_n (r,r',\nu) ={\bf 1} \frac{1}{r}
\delta(r-r')
\ee
For ${\bf M}_n^0$ the Green function is simply by a
diagonal matrix with elements
\be
{\bf G}^0_{n~ii}(r,r',\nu) =
I_{n_i}(\kappa_i r_<)K_{n_i}(\kappa_i r_>)
\ee
where $\kappa = \sqrt{M_i^2 + \nu^2}$. For the Green function
of the operator
${\bf M}_n$ the matrix elements become similarly
\be
{\bf G}_{n~ij}(r,r',\nu) = f_i^{\alpha-}(r<)f_j^{\alpha+}(r_>)
\ee
where the functions $f_i^{\alpha\pm}$ form a fundamental
system of solutions of (\ref{greendgl}) regular as
$r \to 0$ for the minus sign and as $r \to \infty$
for the plus sign.
The correct normalization is obtained by imposing the
boundary conditions
\bea
f_i^{\alpha-}(r) &\simeq&  
\delta_i^{\alpha}I_{n_i} (\kappa_i r) \nonumber \\
f_i^{\alpha+}(r) &\simeq& \delta_i^\alpha K_{n_i} (\kappa_i r)
\eea
as $r \to \infty$. Actually we have solved numerically
the differential equations for the
functions $h_i^{\alpha\pm}$ defined by
\be
f_i^{\alpha \pm} = B_{n_i}(\kappa_ir)
(\delta_i^{\alpha} + h_i^{\alpha}(r))
\ee
where $B_{n_i}$ are the appropriate  Bessel functions. In this
way one keeps track of the free contribution 
$ \propto \delta_i^\alpha$ 
and one has 
\bea
&{\rm Tr}&\left({\bf G}(r,r,\nu)-{\bf G}^0(r,r,\nu)\right)
\nonumber \\
&=& (h_i^{i-}(r)+h_i^{i+}(r)
+h_i^{\alpha-}(r)h_i^{\alpha+}(r))I_{n_i}(\kappa_i r)
K_{n_i}(\kappa_i r)
\eea
to be inserted into (\ref{fnudef}).
The partial wave contributions behave as $n^{-3}$ for large n.
The summation implied by Eq. (\ref{fnudef}) has been performed
up to maximally $\bar n=25$, the asymptotic tail
was appended by fitting
the last five terms with an expression $ a_n =c_3 n^{-3}+c_4 n^{-4}
+ c_5 n^{-5}$ and adding the sum over the $a_n$ from
$\bar n +1$ to $\infty$. The convergence was monitored by
applying this procedure already in each step of
the $n$ summation taking $\bar n = n$.
The convergence was found to be excellent
up to values of $\nu$ of the order 5. It has to be said, though,
that there is considerable cancellation between the
negative $n=0$ contribution and the higher terms. Indeed the
$\nu$ integration over the $n=0$ term alone would be
divergent even after renormalization. This seems to be an
inherent feature for functional determinants for topologically
nontrivial configurations, it is related to the fact
that the centrifugal barriers of the
operator ${\bf M}_n$ at $r=0$ are different from
those of ${\bf M}^0_n$.
This feature makes also a direct application of a theorem
on functional determinants used for the faster and more elegant
method of Ref. \cite{BaKi}  impossible. The deformation of the
centrifugal barriers is not related to our using the singular
gauge for he classical instanton field. In fact it can be shown
by direct calculation that the fluctation equations
do not change under the gauge
gauge transformation (\ref{gaugetr}).

Fortunately the asymptotic behaviour of $F(\nu)$ which is
as $\nu^{-4}$ after renormalization sets already in when
this function has dropped to values of order $10^{-2}$
and there the cancellation is not yet delicate.

There is a further problem which we have to address here which
is related to the coupling of fields with different mass in
the system of gauge, Higgs and Goldstone fields.
While normally the solutions of the coupled system fulfil
vacuum boundary conditions at $r \to \infty$, i. e. the
potential decreases sufficiently fast, the cross terms
${\bf V}_{i4}$ can cause the Higgs field to change the asymptotic
behaviour of the gauge and Goldstone components. The solution
regular at $r = 0$ behaves normally as $\exp (\kappa_i r)$.
If the physical Higgs component is multiplied by ${\bf V}_{i4}$
one obtains a behaviour $\exp( (\kappa_H - M_W)r )$. This expression
enters the right hand sides of the equations for the Goldstone and
gauge fields which themselves behave as $\exp( \kappa_W r)$. So
if $M_H > 2 M_W$ these fields change their asymptotic behaviour.
We find that the integral of the
trace of the Green function over $r$ ceases
to exist. We think that this not a shortcome of the method but
a systematical property of the fluctuation determinant. Indeed
for $M_H > 2M_W$ the Higgs boson can decay into pairs of
gauge particles and also the
singularity structure of perturbative graphs changes qualitatively.
This subjects merits further consideration; here we just restrict
our computation to Higgs masses smaller than $2 M_W$.
The gauge fields cannot, on the other hand, decay into Higgs
particles, since their coupling joins a physical
and a Goldstone Higgs; indeed our coupled system has no
problems of principle for small Higgs masses.


\setcounter{equation}{0}
\section{Renormalization and zero modes}            \label{Renorm}
\par

The Abelian Higgs model is super-renormalizable; all divergences
can be removed by a mass counterterm for the
Higgs field and a counter term for the vacuum loops.
Expanding around $\phi = v$ and using the
corresponding Feynman rules we find divergent tadpole diagrams
of the form represented in Fig. 2, where the internal
lines are the various Higgs, vector and Fadeev-Popov fields.
The various couplings
can be read off from the second order Lagrangean (\ref{lag2}).
For the vertices of the second graph we find  $-3i\lambda/2$ for
the physical Higgs of mass $M_H$, $-i(g^2+\lambda)/2$
for the Goldstone Higgs of mass $M_W$, $ig^2g_{\mu\nu}/2$
for the gauge field and $-ig^2/2$ for the Fadeev-Popov fields.
For the first graph we find the same vertex factors multiplied
by $2v$. As a consequence, in summing the contributions from both
graphs the external line factors combine as
$(\phi-v)^2+2v(\phi-v)=(\phi^2-v^2)$. The contributions from
the gauge field and Fadeev-Popov loops cancel as they should.
The tadpole graphs with external gauge field lines (not
presented in Fig. 1) cancel
against second order graphs as usual in scalar QED.
The counter term action takes the form
\be
S_{ct} = \frac1{2} \delta m^2 \int d^2x (\phi^2-v^2)
\ee
where in unregularized form
\be
\delta m^2 = 3\lambda \int \frac{d^2k}{(2\pi)^2}\frac1{k^2+M_H^2}
+(g^2+\lambda)\int \frac{d^2k}{(2\pi)^2}\frac1{k^2+M_W^2}.
\ee
In the Pauli-Villars
regularization chosen her we rewrite the divergent momentum
integrals via
\bea
\int\frac{d^2k}{(2\pi)^2}&&\left(\frac1{k^2+M_i^2} -
\frac1{k^2+M_i^2+\Lambda^2}\right) \nonumber \\
&&=\int\frac{d^2k}{(2\pi)^2} 2\int_0^\Lambda \nu d\nu 
\frac1{(k^2+M_i^2+\nu^2)^2} \nonumber \\
&&= \frac1{2\pi}\int_0^\Lambda\frac1{M_i^2+\nu^2}.
\eea
so that the divergent terms can be rewritten directly as a
contribution to a counterterm 
$F_{ct}(\nu)$ in the integral over $\nu$.
We find
\be
F_{ct}(\nu) =
\int_0^\infty dr r (f^2(r)-1) \left(\frac{3M_H^2}{\nu^2+M_H^2}
+\frac{M_W^2+M_H^2}{\nu^2+M_W^2} \right)
\ee
If $F_{ct}(\nu)$ is subtracted, $F_{ren}(\nu) = F(\nu)
-F_{ct}(\nu)$  behaves as
$\nu^{-4}$ as $\nu \to \infty$ and the Pauli-Villars
cutoff, i.e. the upper limit of integration,
can be sent to $\infty$.

Instead of subtracting the tadpole contributions from
$F(\nu)$ this subtraction can be performed already in the
partial waves. The tadpole terms can be easily
recognized in the potential given at the end of section 4
as the diagonal terms proportional to $(f^2-1)$. Denoting
these terms by ${\bf V}^{tad}_{ii}$ their contribution
to the first order
Green function becomes
\be
{\bf G}_{n~ii}^{tad}(r,r',\nu)
=\int dr''r'' {\bf G}^0_{n~ii}(r,r'',\nu){\bf V}_{ii}^{tad}(r'')
{\bf G}^0_{n~ii}(r'',r',\nu)
\ee
(no summation over i).
Using some identities for Bessel functions it
can be shown that after taking the trace, integrating over
$r$ and summing up the partial waves one obtains
$F_{ct}(\nu)$. In the actual computation we have removed
the tadpole contributions directly in the partial waves.
As an illustration we show however, in Fig. 3, the function
$F(\nu)$ before the subtraction of the tadpole and Fadeev-Popov
contributions, both of these contributions, and the
final $F^{ren}(\nu)$. It follows from perturbation theory
that the former behave as
$\nu^{-2}$ asymptotically, while the latter behaves as
$\nu^{-4}$. The numerical integration was performed
up to to region where the asymptotic behaviour sets in.
The remaining integral was performed as $\int d\nu \nu^{_-3}$
with a coefficient determined by the last point.
The contribution of the integral from $\nu_{max}$ to
$\infty$ is of the order $.05$ and the error 
introduced by the extrapolation is certainly one order
of magnitude smaller than this value. 

One notes in Fig. 3 that  $F(\nu)$ behaves
for small $\nu$ as $2/\nu^2$, a behaviour that is due to
the translation zero modes and makes the subtraction of
$\ln \epsilon ^2 $ necessary while the lower limit
of the integration is taken to $0$. In practice, the zero
mode pole appears slightly shifted 
to $\nu = \lambda_0 \approx .02$  as can be seen from
the departure of the expected behaviour for $\nu < .1$.
So a term $\ln(\epsilon^2-\lambda_0^2)$ has to be 
subtracted instead. The integrand was, for $\nu < 1$
decomposed into a pole term and a finite
contribution and the former one was integrated analytically.
$\lambda_0$ can be fixed to at least three significant digits
and the finite term turns out to show a smooth behaviour
$\propto \nu$; we think that this procedure introduces
an error of $S_{eff}$ below $.01$. So including the 
estimate for the error in the asymptotic extrapolation
and another $.05$ (i.e. $\simeq 10 \%$) for errors
in the numerical integration 
we think that we have determined $S_{eff}$ to within an
error of $.07$.
 

\setcounter{equation}{0}
\section{Discussion and Conclusion}
\par

The results of our computation of the one-loop
effective action 
\be
S_{eff} = - \lim_{\epsilon \to 0}
(\int_\epsilon^\infty
 d\nu \nu F_{ren}(\nu) + \ln \epsilon^2)
\ee  
are shown in Fig. 4. The
fluctuation prefactor ${\cal D}^{-1/2}$ (including
the counterterm action) is given by
$\exp (-S_{eff}^{ren})$. Due to the subtraction of the
zero mode contribution $\ln \epsilon^2$ it has dimension
$(length)^{-2}$. Since we have used in our computation
the dimensionless variable $M_W r$ the units for the rate
are $M_W^2$ (the action and therefore the zero mode
prefactor being dimensionless). As mentioned in the
previous section we estimate the
error of our numerical result for $S_{eff}$ to be of 
the order of $.07$ units.

In contrast to an analoguos computation of the fluctuation
prefactor for the sphaleron transition in the electroweak theory
here the effects of the quantum fluctuations on the transition
rate remain quite small, less than a factor of $2$. This could 
have been expected on the grounds that the number of 
fluctuating fields small; effectively - in view of the
cancellation between gauge field and Fadeev-Popov degrees
of freedom - we have just the physical and the Goldstone part
of the Higgs field. Also the dimension of space is reduced
from three to two. We think nevertheless that this
expectation had to be checked by a direct computation.

One cannot compare the classical and quantum action without
specifying the dimensionless vacuum expectation
value $v = M_W/g^2$. If $v \simeq 1$ then the classical action
is $ \simeq \pi \times O(1)$. This has to be considered as
an absolute lower limit if one wants to justify the dilute instanton
gas approximation. The fact that the one loop correction 
is then one order of magnitude smaller supports the use
of the semiclassical approximation. It would be interesting to
compare it to lattice simulations.

section*{Acknowledgments}

The authors have pleasure in thanking S. Junker and V. Kiselev
for  discussions.


\newpage
\section*{Figure Captions}

\begin{description}

\item[Fig. 1] The classical action.
We present the action in (dimensionless) units $\pi v^2$.
The full circles are the numerical results, the curve
displays a possible asymptotic
dependence $(M_H/M_W)^(1-\gamma)$,
where $\gamma$ is Eulers constant.

\item[Fig. 2] The divergent tadpole graphs.
The vertex factors and internal line masses are given in the
text. We do not display the analoguous graphs with
external classical gauge field legs since they cancel against
second order contributions.

\item[Fig. 3] The integrand $F(\nu)$ for $M_H=M_W$.
Empty diamonds: $F(\nu)$ before tadpole and Fadeev-Popov subtraction;
empty circles: $F_{ct}(\nu)$; empty squares: the Fadeev Popov
term; full circles: $F_{ren}(\nu)$. The dotted line corresponds
to the behaviour $2/\nu^2$ due to the zero modes. The
dashed line shows the extrapolated asymptotic $\nu^{-4}$ behaviour.

\item[Fig. 4]  The one loop effective action.
The vertical lines are the numerical results for the
bosonic effective action $S_{eff}^{1-loop}$, the length
of the lines indicate the error; the empty
squares are the effective action $S_{eff}^{ferm}$ for
massless fermions given by Eq. (\ref{fermac}). The curves are
spline fits.

\end{description}
\newpage
\section*{Tables}
\begin{table}[h]
 \[
  \begin{array}[t]{|c|c|c|c|} \hline
  M_H/M_W & S_{cl} & S_{eff}^{ferm} & S_{eff}^{1-loop}
  \\ \hline
 0.40& 0.696196  & -0.37853 & 0.315    \\ \hline
 0.60& 0.813053  & -0.38459 & 0.233    \\ \hline
 0.80& 0.912305  & -0.39051 & 0.102    \\ \hline
 1.00& 1.000000  & -0.39616 & -0.025   \\ \hline
 1.25& 1.097914  & -0.40277 & -0.156   \\ \hline
 1.50& 1.186013  & -0.40886 & -0.276   \\ \hline
 1.75& 1.266416  & -0.41445 & -0.379   \\ \hline
 2.00& 1.340550  & -0.41956 & -0.461   \\ \hline
 \end{array}  \]
  \caption{Classical and one-loop actions for various values of
  $M_H/M_W$.
  (left {\it plus} right-handed) fermion. $S_{eff}^{1-loop}$
  is the one loop bosonic action computed here.}
\end{table}

\newpage
\epsfig{file=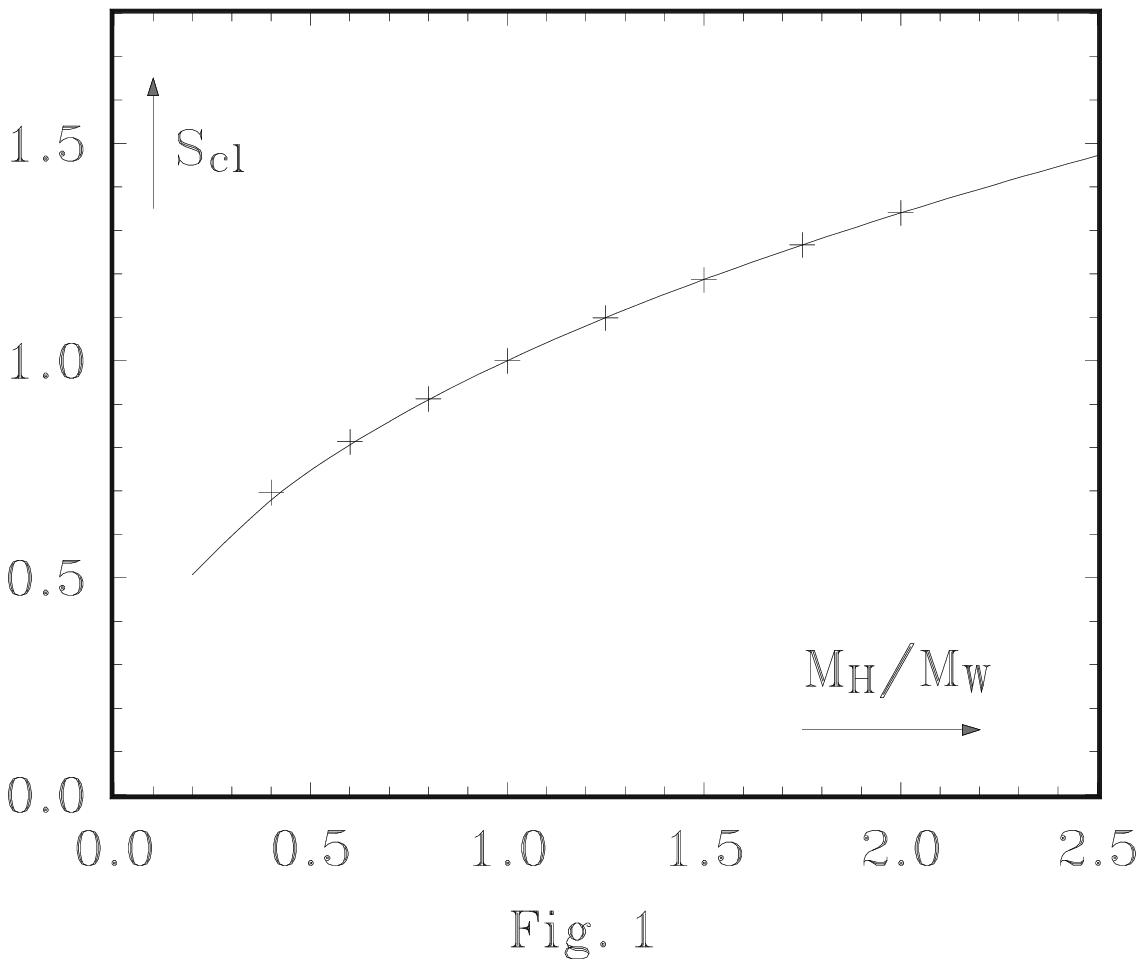,height=9cm}
\epsfig{file=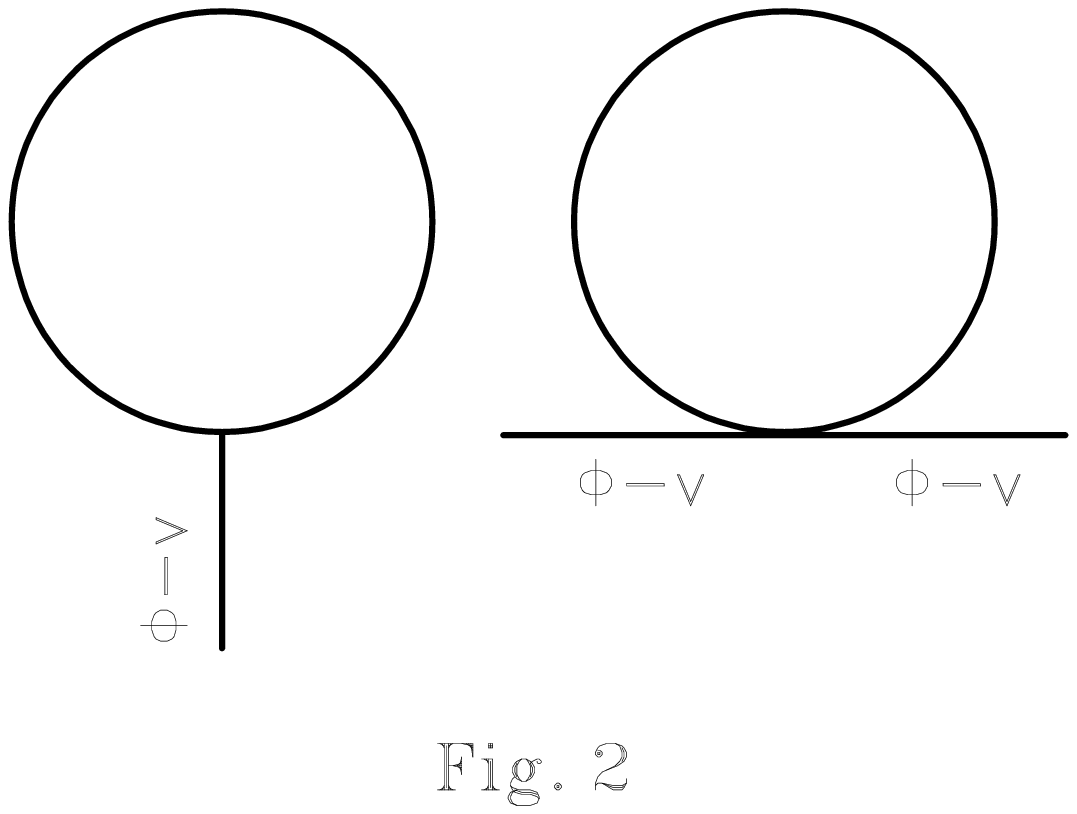,height=9cm}
\epsfig{file=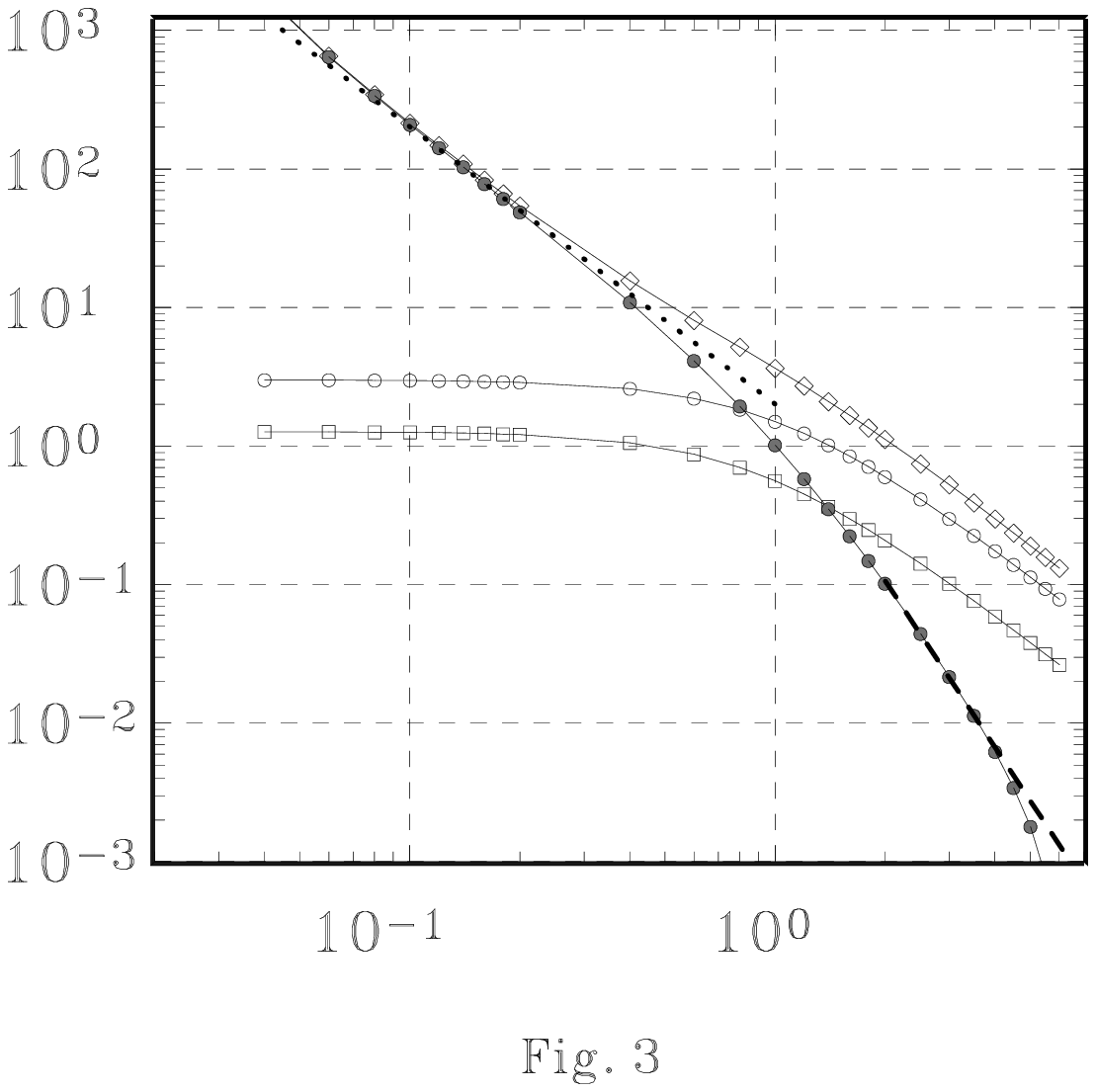,height=9cm}
\epsfig{file=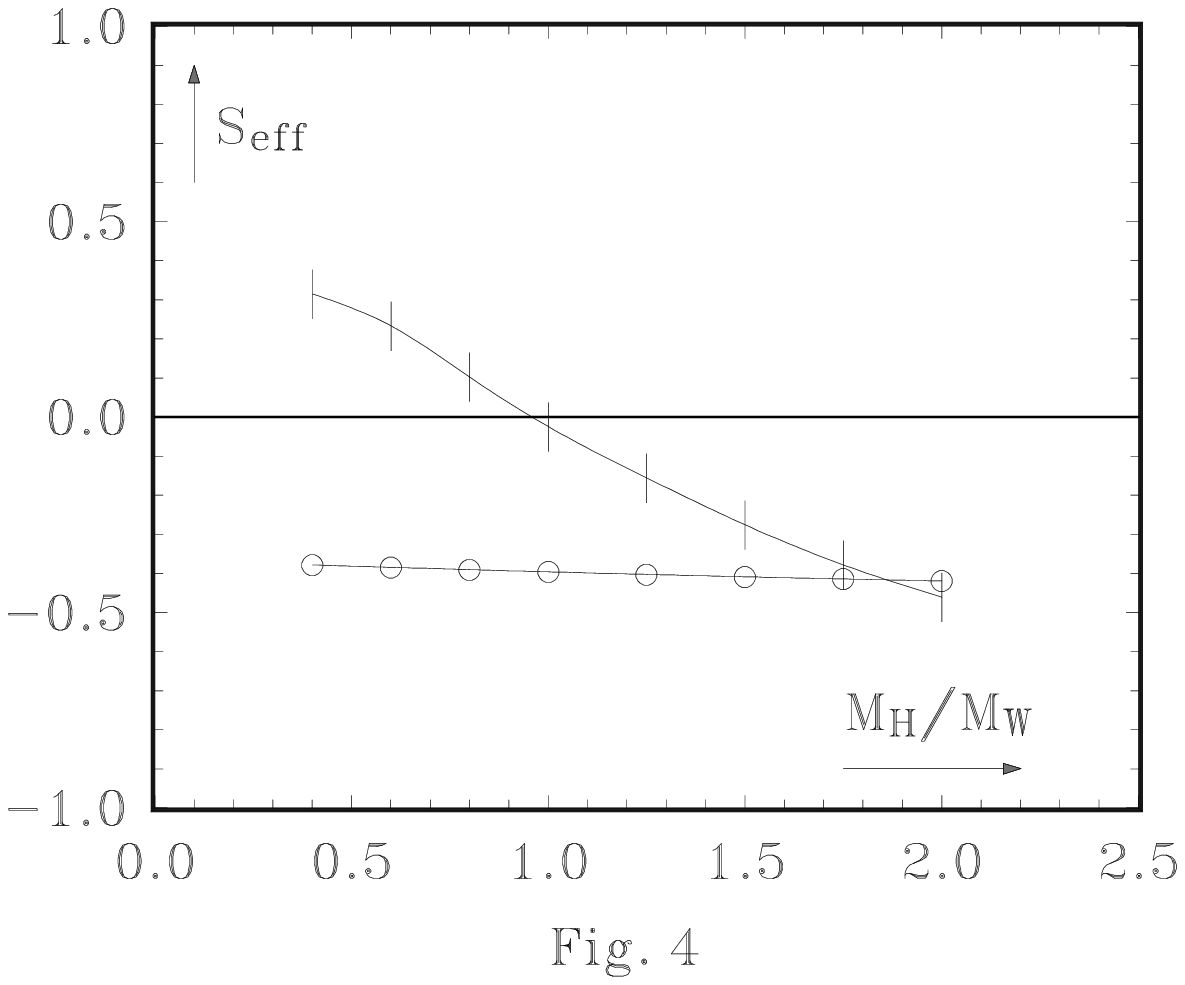,height=9cm}
\end{document}